%% file: main.tex
\documentclass[sigconf,screen]{acmart} 

\AtBeginDocument{%
  \providecommand\BibTeX{{%
    \normalfont B\kern-0.5em{\scshape i\kern-0.25em b}\kern-0.8em\TeX}}}


    
\setcopyright{acmcopyright}
\acmPrice{15.00}
\acmDOI{10.1145/3540250.3549134}
\acmYear{2022}
\copyrightyear{2022}
\acmSubmissionID{fse22main-p559-p}
\acmISBN{978-1-4503-9413-0/22/11}
\acmConference[ESEC/FSE '22]{Proceedings of the 30th ACM Joint European Software Engineering Conference and Symposium on the Foundations of Software Engineering}{November 14--18, 2022}{Singapore, Singapore}
\acmBooktitle{Proceedings of the 30th ACM Joint European Software Engineering Conference and Symposium on the Foundations of Software Engineering (ESEC/FSE '22), November 14--18, 2022, Singapore, Singapore}

\include{macro}

\begin{document}

\title[\toolname: Automating Usage-Based Test Generation from Videos of App Executions]{\toolname: Automating Usage-Based Test Generation\\from Videos of App Executions}

 	\author{Yixue Zhao}
 	\authornote{Both authors contributed equally to the paper}
 \email{yixuezhao@cs.umass.edu}
 \affiliation{%
   \institution{University of Massachusetts Amherst}
   \country{USA}
 }

 \author{Saghar Talebipour}
 \authornotemark[1]
 \email{talebipo@usc.edu}
 \affiliation{
   \institution{University of Southern California}
   \country{USA}
 }

 \author{Kesina Baral}
 \email{kbaral4@gmu.edu}
 \affiliation{
   \institution{George Mason University}
   \country{USA}
 }

 \author{Hyojae Park}
 \email{hyoj.p20@gmail.com}
 \affiliation{
   \institution{Sharon High School}
   \country{USA}
 }
 
  \author{Leon Yee}
 \email{leon.yee000@gmail.com}
 \affiliation{
   \institution{Valley Christian High School}
   \country{USA}
 }
 
   \author{Safwat Ali Khan}
 \email{skhan89@gmu.edu}
 \affiliation{
   \institution{George Mason University}
   \country{USA}
 }
 
    \author{Yuriy Brun}
 \email{brun@cs.umass.edu}
 \affiliation{
   \institution{University of Massachusetts Amherst}
   \country{USA}
 }

 \author{Nenad Medvidovi\'c}
 \email{neno@usc.edu}
 \affiliation{
   \institution{University of Southern California}
   \country{USA}
 }
 
 \author{Kevin Moran}
 \email{kpmoran@gmu.edu}
 \affiliation{
   \institution{George Mason University}
   \country{USA}
 }

\renewcommand{\shortauthors}{Y. Zhao, S. Talebipour, K. Baral, H. Park, L. Yee, S. Khan, Y. Brun, N. Medvidovic, and K. Moran}

\begin{abstract}

Writing and maintaining UI tests for mobile apps is a time-consuming and
tedious task. While decades of research have produced automated approaches
for UI test generation, these approaches typically focus on testing for
crashes or maximizing code coverage. By contrast, recent research has shown
that developers prefer \emph{usage-based tests}, which center around
specific uses of app features, to help support activities such as regression
testing. Very few existing techniques support the generation of such tests,
as doing so requires automating the difficult task of understanding the
semantics of UI screens and user inputs. In this paper, we introduce
\toolname, which automates key steps of generating usage-based tests.
\toolname uses neural models for image understanding to process video
recordings of app uses to synthesize an app-agnostic state-machine encoding
of those uses. Then, \toolname uses this encoding to synthesize test cases
for a new target app. We evaluate \toolname on 374 videos of common uses of
18 popular apps and show that 69\% of the tests \toolname generates
successfully execute the desired usage, and that \toolname's classifiers
outperform the state of the art.

\end{abstract}



\maketitle


%

\input{intro}

\input{appr}
\input{eval}

\input{related-work}
\input{conclusion}


\clearpage
\balance
\bibliographystyle{ACM-Reference-Format}
\bibliography{main}
\end{document}

%% file: macro.tex
\usepackage{setspace}
\usepackage{algorithmic}
\usepackage{graphicx}
\usepackage{textcomp}
\usepackage{xcolor}
\usepackage{soul}
\usepackage[linesnumbered,ruled,vlined]{algorithm2e}
\usepackage{balance}
\usepackage{microtype}
\usepackage{multirow}
\usepackage[normalem]{ulem}
\usepackage{soul}
\usepackage{lipsum}
\usepackage{wrapfig}
\usepackage{verbatim}
\usepackage{tikz}
\usepackage{enumitem}
\usepackage[normalem]{ulem}
\useunder{\uline}{\ul}{}
\usepackage{color,colortbl}
\usepackage{pifont}
\usepackage{tcolorbox}
\usepackage{microtype}

\definecolor{amethyst}{rgb}{0.6, 0.4, 0.8}

\newcommand{\toolname}{\textsc{Avgust}\xspace} 

\newcommand*\circled[1]{\tikz[baseline=(char.base)]{\small{\textbf{
			\node[shape=circle,fill,inner sep=0.75pt] (char) {\textcolor{white}{#1}};}}}}

%% file: intro.tex

\section{Introduction}
\label{sec:intro}




Writing UI tests is time-consuming and tedious. 
The research community has contributed a large body of work that aims to
automatically generate UI tests~\cite{li2006effective, kaasila2012testdroid,
su2017guided,mao2016sapienz, amalfitano2012using, gu2019practicalAPE,
dong2020time, hu2018appflow, lin2019craftdroid, behrang2019atm,
wang2020combodroid}.
Such testing techniques generate a test's inputs, and use a pre-defined 
criterion as the test's oracle. A significant portion of recent 
 work on UI test generation has
focused on \emph{mobile platforms} and has predominantly aimed to 
 discover crashes or maximize code
coverage~\cite{mao2016sapienz, wang2020combodroid, su2017guided,
Moran2016CrashScope, gu2019practicalAPE, dong2020time}. 
However,  studies have repeatedly
found that existing testing techniques in this domain fall short in addressing developers'
needs in
practice~\cite{linares2017developers, Kochar2015Study} or present challenges for
practical adoption~\cite{linares2017developers, haas2021manual}. 

Specifically, mobile developers have a strong preference for
test cases that are closely coupled to app use cases or
features~\cite{linares2017developers}. In line with recent work~\cite{yixue_zhao_fruiter_2020}, we refer to this type of preferred test case as \emph{usage-based} UI test. A usage-based UI test
consists of a sequence of UI
events that mimic \emph{realistic} user behaviors in exercising a specific
feature 
of a given app, such as 
``adding an item to the shopping cart.'' The developer
preference for usage-based tests is due to the fact that such test cases
support specific testing goals in practice, such as regression or performance
testing, which in turn require orientation to common app use
cases~\cite{linares2017developers}.

Automating such testing activities is critical for mobile developers who face
unique challenges related to rapidly evolving
platforms~\cite{Bavota:TSE15,Linares-Vasquez:FSE13}, pressure for frequent
releases~\cite{Hu:EuroSys14,Jones:2014}, and a deluge of feature requests and
bug reports from user
reviews~
\cite{Ciurumelea:SANER'17,DiSorbo:FSE'16,Palomba:ICSE17,Palomba:ICSME15}.
Despite the importance of these usage-based tests for developers, current
automated testing approaches typically do not consider app usages as a goal or 
test adequacy criteria, and as such cannot generate these tests~\cite{linares2017developers, yixue_zhao_fruiter_2020}. 
A growing body of research on the topic of \emph{UI test reuse} (also
sometimes called test migration, test transfer, or test
adaption~\cite{behrang2018test, lin2019craftdroid, behrang2019atm,
mariani2021evolutionary, qin2019testmig}) has begun to explore the
possibility of automating the transfer and adaptation of existing usage-based tests
from a
\emph{source
app} to a behaviorally similar \emph{target app} that contains shared features~\cite{yixue_zhao_fruiter_2020,
hu2018appflow, behrang2018test, lin2019craftdroid, behrang2019atm,
mariani2021evolutionary, mariani2021semantic, qin2019testmig}.
However, these test-reuse techniques have three notable limitations that pose
challenges for developers to adopt
in practice.
\circled{1}~To generate tests for a target app, UI test reuse
requires \emph{pre-existing, manually-written tests} for a corresponding source app. 
In practice, creating these source tests is time-consuming and error-prone, leading many mobile developers to forgo writing them~\cite{linares2017developers, Kochar2015Study}.
\circled{2}~Test-reuse techniques have typically been designed for, and tasked with,
transferring tests between behaviorally similar applications from similar domains (e.g., between two finance or two shopping apps). However, there are many  use cases common across apps
from varying domains (e.g., logging in or changing the theme), which current test techniques would struggle to effectively transfer.
\circled{3}~Many existing techniques rely on expensive and difficult to use
program analyses (e.g., bytecode decompilers, Soot~\cite{soot, vallee2010soot}, Gator~\cite{gator, yang2018static}) that often require access to an app's source code. 
The ease of use and scalability limitations of such underlying utilities have hindered the adoption
of test case transfer tools in practice.

\looseness-1
To help better align automation related to usage-based testing 
with developers needs, we propose \toolname, a technique for \underline{a}pp-\underline{{v}}ideo-based \underline{{g}}eneration of \underline{{us}}age \underline{{t}}ests.
\toolname is a novel
developer-in-the-loop test generation technique that directly addresses the three limitations mentioned above.
\circled{1}~Instead of requiring pre-existing source tests written by domain
experts, \toolname allows for easy creation of source test scenarios through
\emph{screen recordings of app usages}, which are becoming increasingly common
software artifacts for mobile apps~\cite{Cooper:ICSE'21} and can be easily
obtained via crowd workers with no testing expertise.
After video collection and processing, \toolname operates according to two main phases. In the first phase, 
neural computer vision (CV) and natural language processing (NLP) techniques are employed to guide developers through a lightweight screen and GUI widget annotation process for video frames that were automatically identified to contain a touch action. Using this information, \toolname is  able to generate an app-independent intermediate-representation model (\textit{IR Model}), which represents abstract states and transitions of a usage that can be mapped to multiple apps. This procedure is a one-time effort for developers, and once the IR Model is generated, it can be used to generate tests for multiple target apps. \circled{2}~The generality of the  IR Model allows \toolname to synthesize test scenarios across domains, effectively overcoming the second limitation of existing test-case transfer techniques. \toolname's second phase automates the synthesis of new UI test scenarios by guiding a developer with suggestions, made by using predictions from \toolname's  CV and NLP techniques, of which GUI elements must be manipulated to exercise a given app feature or usage. 
\circled{3}~To bolster the applicability and practicality of \toolname from a developer's
perspective, \toolname operates \textit{purely on visual information} encoded into screenshots and video frames from UI-screen recordings. As such, it does not require access to an app's
source code, instrumentation, or expensive program analyses. 
Note that solely relying on app videos as input is a key aspect of \toolname's novelty and it has three major advantages. First, videos are common artifacts that are easily collectible without requiring difficult tool configuration and setup, which are major barriers for adoption~\cite{Kochar2015Study, linares2017developers}. Second, videos can be collected by crowd workers (e.g., real users) with no testing experience, enabling the opportunity to obtain much more training data to cover diverse and realistic usage scenarios across different apps. This can yield more generalized models to generate higher-quality tests. Finally, videos are agnostic to the underlying device and platform, meaning \toolname's design is not tied to Android platforms (where  \toolname is evaluated on), but is applicable to any apps, devices, and platforms (e.g., websites) in principle.

The key research challenge that \toolname tackles is the automated
synthesis of a generalized model of feature usages that can effectively
map test scenarios across apps from a variety of domains, using only screenshots and video frames from screen recordings. The challenge lies in automating two
key tasks: (1)~screen understanding from pixels and (2)~design of an IR
Model that is general enough to capture diverse app usages yet specific enough to allow mapping actions to a given target app for test scenario generation. \toolname accomplishes the
first task through the creation of a bespoke image classification technique, built on top of a neural auto-encoder representation~\cite{Pu:NIPS2016} and BERT-based textual embeddings~\cite{devlin2018bert}. The classification operates at two granularity levels, (i) screen-level, and (ii) GUI widget-level. This classification procedure helps to provide the mapping to our IR Model, and makes use of a rich screen representation obtained by training our neural auto-encoder on the public RICO dataset~\cite{deka2017rico}. \toolname accomplishes the second task by using the information from our classifiers to build a state machine capable of simultaneously capturing {multiple scenarios} from different usages. This yields a richer model of app usage than past test transfer techniques.


In order to build a community resource of usage-based tests, we conducted a user-study to collect
374 video recordings from 18 apps, covering 18 usage scenarios, wherein each usage scenario is exercised by three associated apps, for a total of 54 
unique app-usage pairs~\cite{yixue_zhao_fruiter_2020,
hu2018appflow}. 
Using this data, we conducted an empirical evaluation to measure the efficacy of \toolname in generating usage-based tests that closely mirror those created by humans during our user study.  First, we examined \toolname's test generation capability by simulating a developer interacting with \toolname's suggestions, and measured how closely generated tests matched analogous tests created by human users. Next, to gain a better understanding of \toolname's performance during test generation, we evaluated \toolname's classifiers compared to state-of-the-art techniques. Our results show that \toolname is able to generate tests that effectively exercise target-app features and closely match human tests in terms of the screens visited and actions performed. Additionally, \toolname's classifiers significantly outperform state-of-the-art techniques and show promising performance for our generated developer-in-the-loop recommendations. 

We have developed \toolname with open science in mind, making it both practical
to use for developers, and easily reusable by
researchers to foster future research in this area. We make publicly available all of our source code, trained models, and annotated evaluation data collected during our user study~\cite{ourRepo}. 
\toolname's pipeline can be
easily adapted to create various usage models of interest by simply changing
the input videos, such as including additional
usages and apps. 
As such, \toolname not only lays a foundation for future work on 
usage-based test generation, but also represents a living repository for the
software engineering community to study related problems.

In summary, this paper makes the following contributions:
\begin{enumerate}
\item We introduce \toolname, the first
technique capable of generating \emph{usage-based} tests by learning from
{app
videos}.
\item We develop a novel image classification technique to translate
app videos into an app-independent intermediate representation based on
{vision-only} information, which  largely outperforms the
state-of-the-art.
\item We implement a reusable pipeline to train IR models based on app
videos
that can be applied to various downstream tasks, and further provide 125
pre-trained models that can be used by developers directly.
\item We collect 374 app videos and conduct an empirical evaluation to demonstrate
\toolname's
effectiveness in assisting developers with generating usage-based tests. 
\item We provide a public repository~\cite{ourRepo} that contains \toolname's
artifacts to foster future research, including \toolname's source code, our
pre-trained models, labeled datasets, benchmarks used, and their
corresponding
results. 
\end{enumerate}

%% file: appr.tex
\section{The \toolname Approach}
\label{sec:appr}

\toolname
is an automated approach that aims to assist developers with the generation
of usage-based tests to mimic realistic usage scenarios. 
\toolname operates in three
phases. (1)~It processes recorded videos of different apps' usages by
applying neural CV and NLP to detect user actions in individual video frames.
(2)~\toolname uses this information to generate an app-independent state
machine-based IR Model. (3)~Finally, \toolname leverages the IR Model to
generate tests for a new (i.e., ``target'') app. In this section, we provide
an overview of \toolname's workflow, and then detail its three phases.
\begin{figure*}[t!]
	\includegraphics[width=
.95\textwidth]{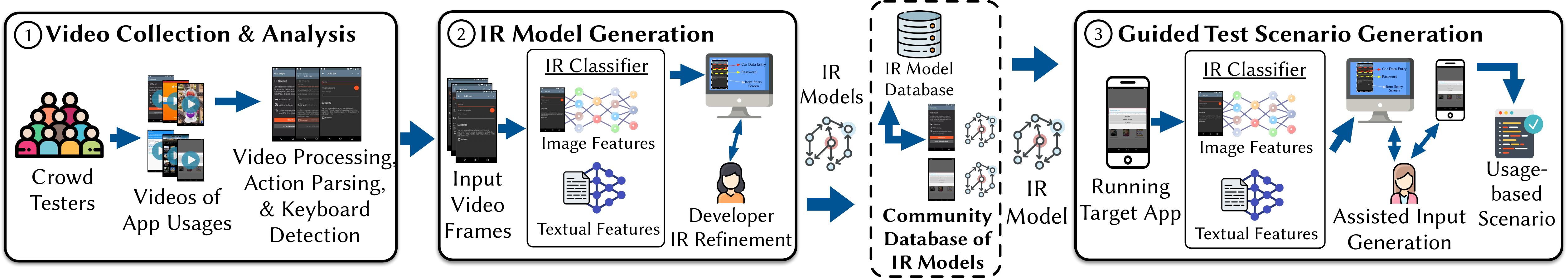}
\vspace{-1em}
\caption{\toolname's three-phase workflow.}
\vspace{-1em}
	\label{fig:design}
\end{figure*}

\subsection{\toolname Overview}

\toolname functions as a human-in-the-loop  tool to provide suggestions of input events for developers in the creation of usage-based tests. 
This design decision is guided by the nature of \emph{usage-based} tests, since each usage scenario may have various correct ways of being tested. For instance, there may be different ways to execute the login scenario in an app, such as logging in using username and password or by using user's existing social media accounts. Thus, providing suggestions to a developer allows for flexibility in generating tests that are tailored to a given app usage and testing objective.
Figure~\ref{fig:design} depicts
\toolname's workflow, which consists of three principal phases:
\ding{172}~\textit{~Video Collection \& Analysis}, \ding{173}~\textit{IR
Model Generation}, and \ding{174}~\textit{Guided Test Scenario Generation}.

During the \textit{Video Collection \& Analysis} phase, crowdsourced workers are tasked with collecting videos of app usages. These videos are then analyzed in a fully automated process that involves deconstructing the video into constituent frames, identifying touch-based actions that were performed on the an app's UI (which builds upon past work in app-video analysis~\cite{bernal2020translatingv2s}), and eliminating sensitive information such as user passwords. 

Next, in the \textit{IR Model Generation} phase, \toolname assists a developer 
 with labeling screens and individual GUI widgets from processed video
frames into categories, which \toolname can then use to generate an
app-independent IR Model. This is a semi-automated process wherein a
developer is presented with a screen and \toolname provides top-k suggestions
for the labels that should be applied to both screens an exercised GUI
widgets. These suggestions are made using a combination of visual- and
text-based classifiers that operate upon the video frames extracted in
the prior phase. After the labels have been applied by the developer,
\toolname is able to automatically generate the state machine-based IR Model
for the usage, merging it with other similar usages in a shared database.
This phase is intended to be a one-time cost, wherein developers contribute
their crowdsourced IR Models of various app usages to a collective community
database for future use.

Finally, in the \textit{Guided Test Scenario Generation} phase, \toolname
assists developers by providing top-k recommendations for actions that should
be performed on given screens of a previously unseen target app in order to
exercise a specified app feature (e.g., adding an item to the shopping cart).
This process starts from the initial screen of the app and runs until the
specified feature is exercised. This functions similarly to the IR Model
generation phase, but in reverse order: the model is used in conjunction with
\toolname's classification techniques to recommend event inputs to 
developers.

\subsection{\textbf{Video Collection \& Analysis}}
\label{sec:appr:video analysis}

Given a set of collected videos of app usages, \toolname's video analysis
processes them into frames that serve as the inputs for \toolname'
AI-assisted IR Model generation (Section~\ref{sec:appr:IR Model generation}).
Specifically, \toolname first identifies the user actions in the videos and
extracts their corresponding \emph{event frames}, which are the key video
frames that capture the user interactions via the \emph{touch
indicator}.\footnote{\toolname requires enabling the display of the
\textit{touch indicator}, which can easily be done in, both, the Android and
iOS settings menus, even by inexperienced users.} An example of {event frame}
is shown in Figure~\ref{fig:frame2model}, where the {touch indicator} points
to the user interacting with the ``app menu'' button in the top-left corner.
As a final step, \toolname filters the extracted event frames by eliminating the frames that contain sensitive user information. 

\subsubsection{Action Identification \& Event Frame Extraction}

\toolname builds upon the analyses introduced by
V2S~\cite{bernal2020translatingv2s, havranek2021v2stool} to identify user
actions and event frames. V2S is a recent technique that leverages neural
object detection and image classification to identify the user actions in a
video, and automatically translates these actions into a replayable scenario.
We extended V2S to work with GPU clusters, to enable it to process large
numbers of videos in parallel.
The outputs of our extended video processing technique are (1) a sequence of
\emph{event frames} of a given video and their associated user actions (i.e.,
click, long tap, and swipe); and (2) the \emph{coordinates} of the touch
indicator in each event frame~\cite{Lee2018JunCoordinates}.

\subsubsection{Event Frame Filtering}

\toolname filters the extracted event frames by eliminating the frames that are associated with the \emph{typing} action, since they may expose user's private information such as password. 
Note that \toolname only eliminates the frames where the user types the text content on a keyboard, but still keeps the frames where the user selects which input field she intends to enter the text content. For example, the sequence of video frames related to ``typing user password'' consists of (1) a frame associated with clicking the password field, and (2) a group of consecutive frames associated with typing each individual character in the password. \toolname only eliminates the latter, while maintaining the former to represent ``typing user password'' action in the usage scenario.

\begin{figure}[b!]
\vspace{-1em}
	\includegraphics[width=0.35\textwidth]{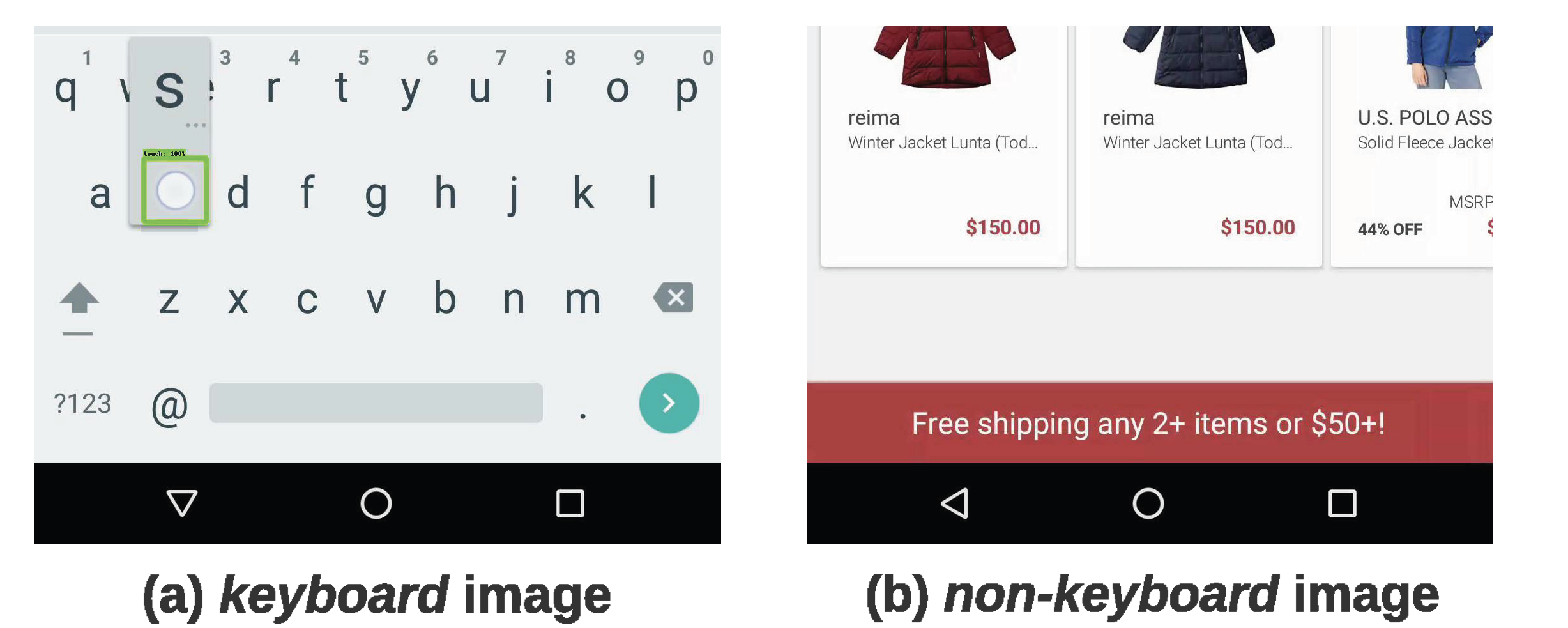}
	\vspace{-1em}
\caption{Examples of the training data used in \toolname's keyboard
classifier during the event frame filtering.}
	\label{fig:keyboard}
\end{figure}

To do so automatically, we trained a binary image
classifier to recognize whether an event frame contains a keyboard image. Our
classifier is based on a CNN architecture with 4 blocks, each consisting of a
Convolution, a BatchNorm, a ReLU, and a Dropout
layer~\cite{Verma2021MaybinaryClassifier}.
The CNN is trained on cropped screenshots that depict the area of the screen
where keyboard {may} appear, since this area is standard for mobile devices,
as shown in Figure~\ref{fig:keyboard}. We decided to focus on the region of
the screen where the keyboard appears, as opposed to the entirety of the
screen, based on empirical evidence collected while tuning our classifier, as
the former setting dramatically improved the classifier's accuracy. We
sourced \emph{non-keyboard} training data by randomly selecting 4,926 app
screenshots without the keyboard from the publicly available RICO
dataset~\cite{deka2017rico}. 
The {non-keyboard} training data do not contain any subject apps used in our
evaluation. Because screens that display a keyboard cannot be automatically
identified using the GUI metadata provided in the RICO dataset, we
additionally sourced 5,605 \emph{keyboard} training data images from the
video frames in the dataset collected for \toolname's evaluation. Our
training data relies on standard Android keyboard images, but can be easily
extended to additional keyboard types.

Next, the cropped screen region where a keyboard may appear for each event
frame is fed to the trained keyboard classifier, and will be classified as
either a \emph{keyboard} or \emph{non-keyboard} frame. For the keyboard
frames, \toolname further verifies whether the associated action is
\emph{typing}, based on whether the touch indicator falls in the keyboard
region. This is determined by the touch indicator's {coordinates} on the
screen~\cite{Lee2018JunCoordinates}, which are obtained from V2S. In the end,
the event frames that contain a {typing} action are eliminated.

\looseness-1
Note that this frame filtering process not only addresses the privacy issue
as discussed earlier, it also largely reduces the number of event frames used
to represent an app's usage. For example, a two-minute sign-in video from the
app 6pm contained over 3,000 video frames originally, but only 8
\emph{filtered} event frames. These 8 frames are sufficient to represent all
relevant user actions without the duplicated or privacy-exposing frames in
the original video frames.

\subsection{AI-Assisted IR Model Generation}
\label{sec:appr:IR Model generation}

\begin{figure}[b!]
  \includegraphics[width=0.475\textwidth]{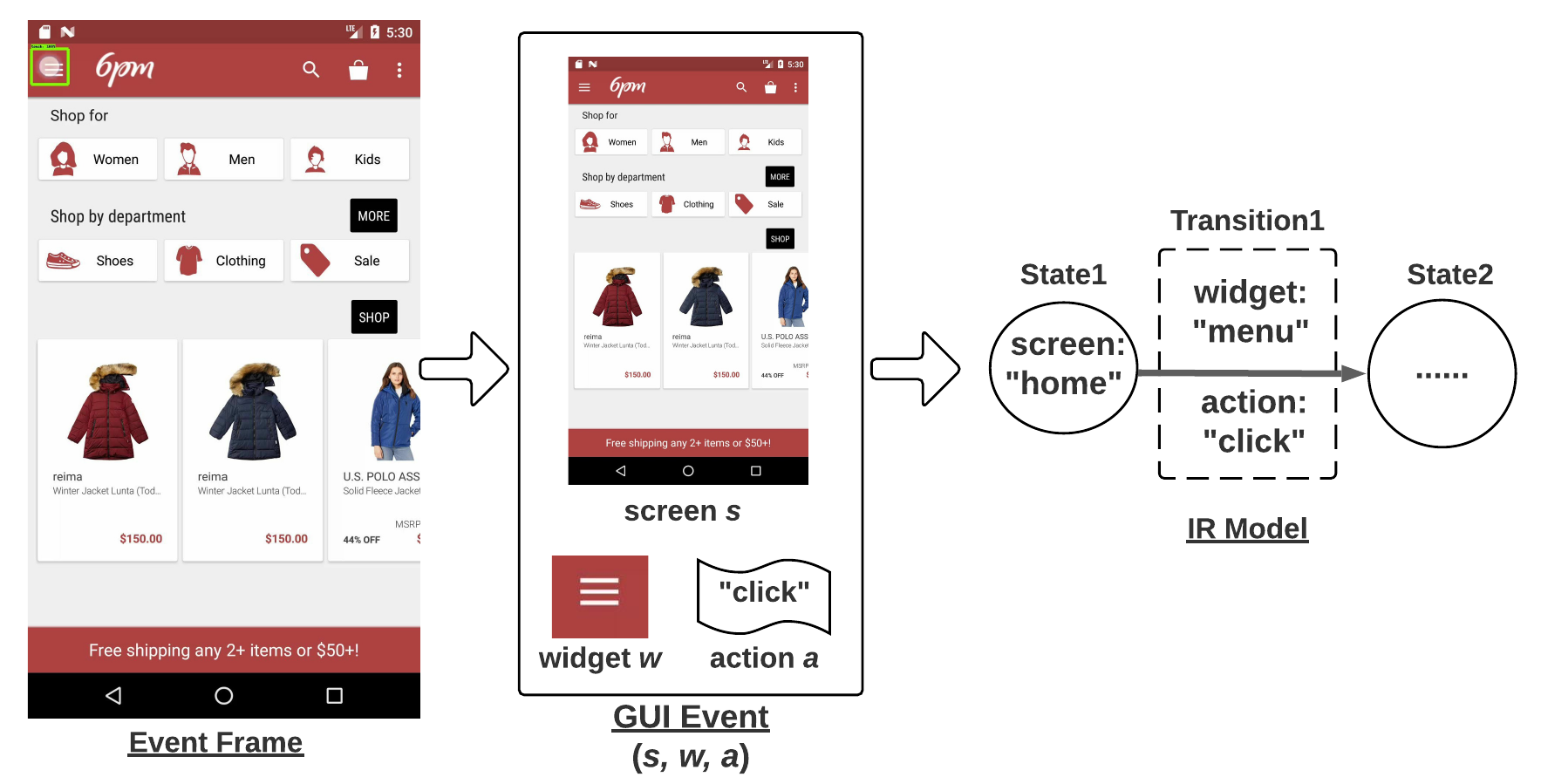}
  \center
  \vspace{-1.5em}
  \caption{An example of converting a 6pm's Event Frame into a GUI Event
  triple, and an app-independent IR Model.}
  \label{fig:frame2model}
  \vspace{-1.5em}
\end{figure}

\toolname uses the filtered event frames from the previous phase as inputs
and translates them into app-independent IR Models of app usages. The key
technical challenge in this phase stems from \toolname's use of {video}
inputs, which forces us to rely solely on {visual information} encoded into
the pixels of the video frames. We address this challenge in two steps: (1)
we break down event frames into GUI events and (2) use image classification
techniques to assist developers in translating GUI events to their
corresponding IR Models. As an illustration, Figure~\ref{fig:frame2model}
demonstrates the key artifacts in this process using a single event frame
extracted from a popular shopping app 6pm as an example. We now detail these
two steps.


\subsubsection{Transforming Event Frames to GUI Events}
\label{sec:appr:gui_event}

We define a \emph{GUI event} as a triple $({{s}},{{w}},{{a}})$, where ${{s}}$
is the app {screen} that shows a snapshot of the app's execution state;
${{w}}$ is the GUI {widget} the user interacts with; and ${{a}}$ is the
corresponding {action} the user performs, such as \texttt{click} or
\texttt{swipe}. The GUI widget \emph{w} is optional since certain actions
(e.g., \texttt{swipe}) are not associated with any widgets. \toolname
converts event frames into GUI event triples in a two-step process: widget
extraction and action identification.

\vspace{1.5mm}
\noindent \textbf{{Widget Extraction.}}~~
Extracting individual widgets from a screen presents two major challenges.
First, each widget's {bounding box} must be identified without reliance on
source code-level information that is available on platforms like
Android~\cite{androidUILayouts}.
Second, \toolname needs to isolate the precise bounding box of the widget
with which the user is interacting, such as the app-menu button in the
top-left corner of the screen in Figure~\ref{fig:frame2model}.

%

\begin{wrapfigure}{r}{0.41\columnwidth}
	\includegraphics[width=0.2\textwidth]{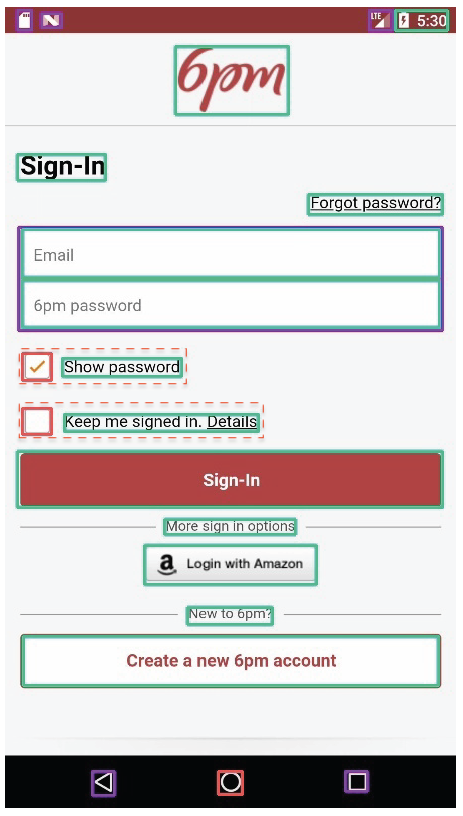}
\caption{\toolname adjusts the GUI-element bounding boxes detected by
UIED, depicted by the two dashed rectangles.}
	\label{fig:UIED}
\end{wrapfigure}

To detect the bounding boxes of GUI widgets, we modified
UIED~\cite{xie2020uiedTool, chen2020objectUIED}, a state-of-the-art 
tool that combines unsupervised CV and deep learning, and
applied it on the {screens} extracted from \toolname's previous phase. Given
an input screen, UIED detects textual and visual GUI elements and produces
their bounding boxes, as depicted with solid rectangles in Figure~\ref{fig:UIED}.
However, UIED treats each visual and textual GUI element separately,
which can lose important semantic
information. 
For example, if the {touch} indicator refers to a checkbox, the
corresponding GUI element detected by UIED in Figure~\ref{fig:UIED} would
be one of the two checkboxes only,
leaving unclear whether the extracted widget is intended
to be associated with ``Show password'' or ``Keep me signed in''. 

To remedy this, we modified UIED to group the {visual} GUI elements together
with their surrounding {textual} elements, if any. \toolname iterates through
all {visual} elements detected by UIED and identifies their closest GUI
elements. If the closest GUI element is both a {textual} element and is in
the same line as the visual element---defined as being vertically collocated
based on a customizable threshold---then the bounding box of the visual
element will be updated to include the textual element as well. In
Figure~\ref{fig:UIED}, this results in the two checkboxes being grouped with
their corresponding labels, as depicted by the dashed rectangles.

Next, the detected GUI elements' bounding boxes 
are used by \toolname to automatically crop out the widget ${{w}}$
to which the touch indicator refers. 
\toolname combines the widgets detected by its modified UIED with the coordinates of the touch indicator obtained from the modified V2S (recall Section~\ref{sec:appr:video analysis}) to identify all candidate widgets for cropping, covering three possible cases: (1)~The simplest case is when only one widget's bounding box
covers the touch indicator, in which case \toolname  crops that widget as-is. 
(2)~If no widgets' bounding boxes cover the touch indicator, \toolname 
repeatedly expands each widget's bounding box based on a customizable threshold, 
until a suitable widget is found. \toolname's default threshold is set at 10 pixels. (3)~When multiple 
widget candidates are found, \toolname first eliminates the ``coarse-grained'' candidates whose boundaries completely 
cover any of the other candidates (e.g., ``sign-in form'' that covers ``username'' widget), and then selects the widget whose center point is closest to 
the touch indicator's coordinates. 

\vspace{1.5mm}
\noindent \textbf{{Action Identification.}}~~ 
To identify the action ${{a}}$ in the GUI event triple, \toolname leverages V2S's action identification procedure, which analyzes the coordinates of touches detected in consecutive video frames and classifies actions according to a set of heuristics~\cite{bernal2020translatingv2s, havranek2021v2stool}. 
V2S is able to identify 
{clicks}, {long taps}, and {swipes}. We reused V2S's heuristics for
{click} and {long tap}, and extended its swipe detection heuristic to additionally detect
the \textit{direction} of the swipe.

\subsubsection{Transforming GUI Events to IR Models}
\label{sec:appr:ir_model}

\toolname's IR Model generation is 
a developer-in-the-loop process. This section explains
how \toolname  provides recommendations to assist developers in
translating GUI events into their app-independent
IR representations (recall the example in Figure~\ref{fig:frame2model}).

\begin{figure}
  \includegraphics[width=0.925\columnwidth]{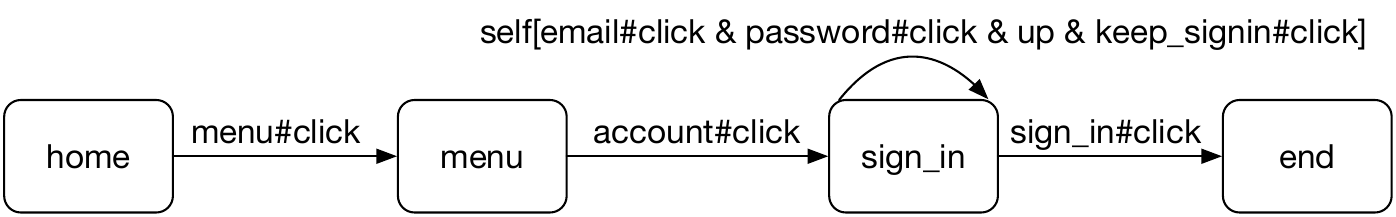}
  \vspace{-1em}
  \caption{The IR Model generated from a sign-in video collected from the app 6pm.}
  \label{fig:ir_model}
\vspace{-1.5em}
\end{figure}

\toolname's IR Model is defined as a finite state machine (FSM) that captures
app usages. Figure~\ref{fig:ir_model} shows an example IR Model converted
from one of 6pm's sign-in videos. 
Each state in the IR Model represents a particular app screen and is captured as an app-independent
\emph{canonical screen}, while
each {transition} 
represents a user interaction with a \emph{canonical widget} and its
corresponding action. A self-transition (e.g., shown in the ``sign\_in'' state in
Figure~\ref{fig:ir_model}) means that the app stays on the same screen during
certain user interactions.


The key challenge in translating a GUI event triple $(s,w,a)$
into \toolname's IR Model is to properly abstract away and capture in an app-independent 
manner the app-specific {screens} $s$ and {widgets} $w$. Each {action} $a$ in the GUI event
triple is translated as-is, including {click}, {long tap}, and swipe
up/down/left/right. With the translated canonical screens, canonical widgets,
and actions, the final IR Model can be constructed by iterating through the
sequence of GUI events of a particular usage. 

We formulate the translation of screens and widgets as a classification
problem, where app-specific screens and widgets are classified into their
canonical counterparts (categories) that are shared across different apps. 
To this end, we build upon and extend
app-independent categories defined by previous work~\cite{hu2018appflow,
yixue_zhao_fruiter_2020}, resulting in 37 canonical screens and 74
canonical widgets. Example canonical screens are ``home screen'', ``password
assistant page'', and ``shopping cart page''. Example canonical widgets are
``account'', ``help'', and ``buy''. The complete sets of canonical
screens and widgets are available~\cite{ourRepo}. Note that these categories are not directly tied to our subject apps used in the evaluation, but can generalize across diverse apps. 
To facilitate the extensibility of new canonical screens/widgets, we have created a data labeling tool using Label Studio~\cite{labelstudio}, allowing future research to further tailor and improve the canonical categories for both screens and widgets.
%

We note that our classification problem provides a unique challenge
as it relies only on {information} from {app screenshots}.
There are no existing techniques in the mobile-app domain that have previously 
addressed this problem~\cite{hu2018appflow, deka2017rico, li2021screen2vec} in a context similar to ours.
Moran~et.al.~\cite{moran2018REDRAW} were one of the first to use a CNN for
widget classification from GUI component images. However, their classifier
only functioned on 15 general widget categories. Our larger number of 74 categories represents a more challenging classification problem that requires the use of both textual and visual features to achieve reasonable accuracy. Another recent technique for
screen classification in this domain is Screen2Vec
(S2V)~\cite{li2021screen2vec}, which aims to produce embeddings for app
screens that can be used for downstream classification tasks, such as ours.
However, S2V cannot be applied on screenshots alone as it requires the UI
layout information of an app screen that is specific to the
Android platform~\cite{androidUILayouts}. Obtaining such information requires extraction
using third-party tools (e.g.,
Appium~\cite{appium}, UIAutomator~\cite{uiautomator}), and would 
sacrifice the practicality of usage collection through videos. 


\vspace{1.5mm}
\noindent \textbf{Screen Classifier.}~~
To classify a given app-screen image into its canonical screen, \toolname leverages both {visual} and {textual} features, wherein textual features are extracted from the image using the Tesseract OCR engine~\cite{smith2007Tesseract}. More precisely, we make use of a pre-trained autoencoder model to encode the screen's visual information, and a pre-trained BERT language model~\cite{devlin2018bert} to encode the screen's textual information. \toolname uses a three-layer convolutional autoencoder with max pooling~\cite{Goodfellow-et-al-2016} and is tasked with encoding an image into a high dimensional vector space, and then decoding the image vector to reconstruct the original image, hence employing a self-supervised training process. 

\begin{figure}[t!]
	\includegraphics[width=0.35\textwidth]{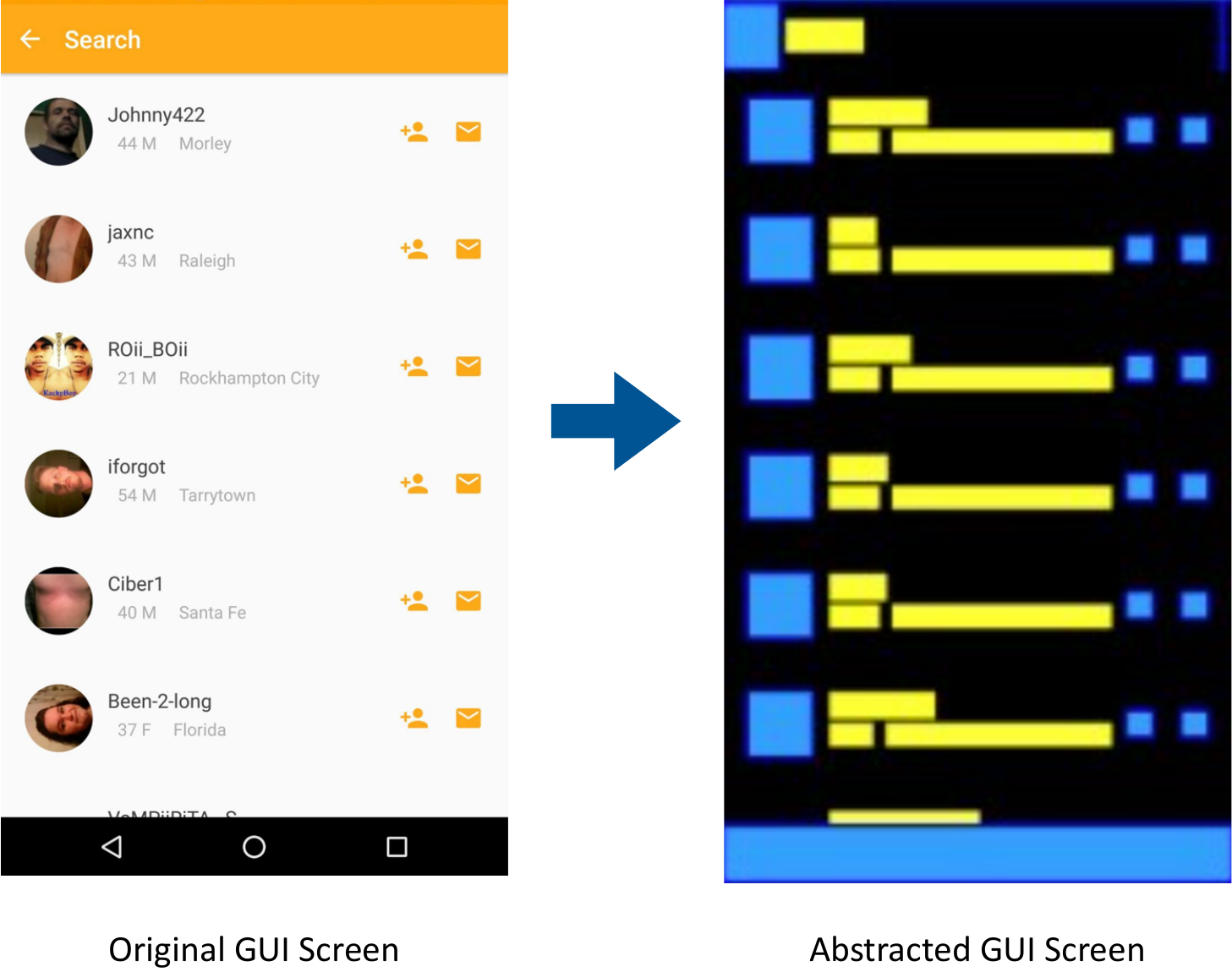}
\vspace{-1em}
\caption{\toolname's screen abstraction process.}
\vspace{-1.5em}
	\label{fig:abstract}
\end{figure}

\looseness-1
As past work has shown~\cite{Li:ASE2019}, learning features or patterns
directly from the pixels of UI screens can be difficult due to the
variability in GUI designs across apps. Therefore, to train and use our
auto-encoder to learn \textit{app-agnostic} visual patterns, we
re-implemented the screen segmentation approach introduced by
REMAUI~\cite{nguyen2015REMAUI}, and use the segments to generate
{abstracted} versions of screens from the RICO dataset~\cite{deka2017rico}. As illustrated in Figure~\ref{fig:abstract}, 
in these abstracted screens text components are transformed into yellow boxes and non-text components
into blue boxes, on a black background. We trained
\toolname's autoencoder on 33,000 abstracted images from the RICO
dataset~\cite{deka2017rico}, and to classify an incoming
screen, we run it through this abstraction process, and then through the
{encoder} of our autoencoder network to extract the feature vector.

\toolname's screen classifier leverages linear layers to combine the
autoencoder and BERT embeddings and classify the screens. The architecture
for the screen classifier consists of three blocks, each containing a linear
layer, BatchNorm, a ReLU activation function, and a dropout layer. These
blocks are followed by a fully connected output layer that applies softmax
function to predict the probability distribution of different screen classes.
We then train the screen classifier on partitions of data collected for our
evaluation, introduced in Section~\ref{sec:eval:setup}, where individual
classifiers for each app were trained on data sourced from other apps. This
process produced 18 pre-trained screen classifiers for each of our subject
apps, and will be reused in \toolname's test generation phase (see
Section~\ref{sec:appr:guided teset generation}).

\begin{figure*}[t]
\includegraphics[width=\textwidth]{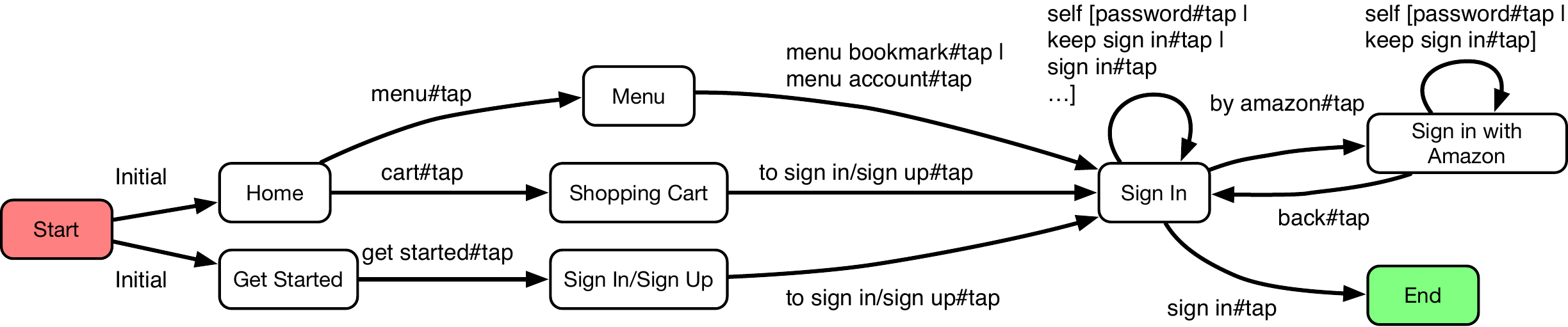}
\caption{A simplified example of the merged IR Model learned from three sign-in videos of  the 6pm and Etsy apps.}
\label{fig:merged_model}
\end{figure*}

\vspace{1.5mm}
\noindent \textbf{Widget Classifier.}~~
To classify a given app-widget image, \toolname leverages its textual, visual, contextual,  type, and spatial information:
(1)~the widget's text is extracted from the widget's
image using Tesseract~\cite{smith2007Tesseract} and  then
encoded using the pre-trained BERT model; (2)~the canonical screen of the screen
image to which the widget belongs is mapped to an \texttt{id} and transformed into a
continuous vector via an embedding layer; (3)~the visual features of the widget are encoded with the
pre-trained ResNet model~\cite{he2016deep}, widely used for encoding
images; (4)~the UI widget class type (e.g., \texttt{\small EditText},
\texttt{\small ImageButton}) is obtained 
using the
classification method introduced by ReDraw~\cite{moran2018REDRAW} and refined by S2V~\cite{li2021screen2vec}, then mapped by
an embedding layer 
into a continuous vector; and (5)~the widget's location on the screen is
obtained by dividing the screen into 9 zones and then transformed to a
continuous vector via an embedding layer.

\toolname's widget classifier then adapts a similar architecture to its
screen classifier---three blocks of linear layers followed by a fully connected
output layer applying softmax---to combine the different feature vectors
discussed above. Note that the embedding layers for the canonical screen,
widget location, and widget class type features are optimized during the
training phase of the widget classifier to generate meaningful embeddings for
each of these input features.

Similarly to the screen classifier, \toolname's widget classifier is trained on
our dataset (Section~\ref{sec:eval:setup}), and produces 18 pre-trained
models that are reused in \toolname's test generation phase.


\toolname's screen and widget classifiers are able to provide the standard top-k labels with different confidence levels. 
The top-k labels are then recommended to developers in labeling the screens and widgets and, in turn,
  the IR Models are constructed using the specific labels selected by developers.
Note that the IR labels refined by developers and the generated IR Models can
be reused by future work. We have thus created a database~\cite{ourRepo} to serve as a
living repository for this problem domain, as also depicted in Figure~\ref{fig:design}.

\vspace{-2mm}
\subsection{Guided Test Scenario Generation}
\label{sec:appr:guided teset generation}

%


\toolname assists developers in generating usage-based tests for their (``target'') apps by leveraging the IR models described above.
Given a usage of interest, \toolname selects the relevant IR Model(s) from the IR Model Database (recall Figure~\ref{fig:design}), and uses them to guide the test generation.
Internally, IR Models of the same usage are represented as a single merged model where multiple scenarios for a given usage populate the same state machine. 
Specifically, the merged model is constructed by using the union of all the edges from all the IR Models of the same usage, demonstrating ``all possible transitions''.
A simplified example of the merged IR model for the sign-in usage is shown in Figure~\ref{fig:merged_model}. 
This unified model of scenarios for a single usage gives \toolname the ability to generate multiple test scenarios for a target usage on an unseen app. \toolname's test scenario generation phase has three principal components: (1) State Extractor, (2) State Matcher, and (3) Event Generator.
We first describe the test generation workflow, and then discuss the three components.

\looseness-1 \toolname's test scenario generation phase is an iterative process that continues to generate the test inputs based on the target app's current state, until the end condition is met (i.e., the target feature is executed).
The process begins by launching the target app and running \toolname's Screen Classifier to retrieve the most likely canonical categories of the target app's starting screen.
\toolname then presents the developer with these top-k classification results, and the  canonical category selected by the developer is  used to match the target app's current device screen to the canonical screen states in the IR Model.
Next, \toolname  recommends the top-k app widgets for developers to interact with by using its Widget Classifier 
to map the {canonical widgets} in the IR Model to the widgets on the target app's current  screen. \toolname then checks whether the test should complete based on whether the  widget chosen by the developer will lead to the end state in the IR Model.
If not, the chosen widget is  triggered, the target app's next state becomes its current state, and this process repeats until the end state in the unified IR Model for a given usage is reached.

\subsubsection{{State Extractor}}
\label{sec:appr:state_extractor}

\toolname's State Extractor extends recent work on the MAPIT~\cite{Talebipour2021MAPIT} test case transfer tool. 
Specifically, for a given target app \toolname extracts (1)~the bitmap of the current screen, (2)~the graph representation of the app screen's UI layout hierarchy~\cite{androidUILayouts}, and (3)~the boundaries of each UI widget and their corresponding cropped images.
The UI layout hierarchy is an XML file that contains the information of all the UI widgets on the target app's current screen, such as their position, size, textual attributes (e.g., ``Sign In''), and class name (e.g., \texttt{ImageButton}). 
The extracted information is  used by \toolname to generate tests, and also to explore different variants of \toolname's classifiers as discussed in Section~\ref{sec:eval}.

\subsubsection{{State Matcher}}
\label{sec:appr:state_matcher}

As discussed previously, \toolname uses its Screen Classifier to suggest the top-k candidates for the {canonical category} of a  target app's given screen. Once the developer  selects from one of the suggested categories,  \toolname maps the current screen to the corresponding state in the IR Model. 
Since all possible transitions captured in the IR Models are known, 
\toolname is able to recommend the target app's widget(s) with which the developer should interact by using a combination of the Widget Classifier (recall Section~\ref{sec:appr:ir_model}), information obtained from the State Extractor (recall Section~\ref{sec:appr:state_extractor}), and a set of pre-defined heuristics. 

\looseness-1
The number of widgets  to be recommended by \toolname is determined by a configurable threshold. \toolname first checks whether the target widgets match the expected canonical widgets from the IR Model based on a set of heuristics that can be divided into two  categories. The first category are heuristics that infer a widget's type based on the UI class of its parent widget. This allows \toolname to bypass the noise that may be present in the data associated with an individual widget. For example, \toolname identifies a widget that represents a menu item, not by trying to capture all possible menu items, but much more simply by comparing its parent widget's UI class to \emph{ListView}. The second category are heuristics that correlate the textual data of a widget with similar terms associated with  each of the canonical widgets (e.g., the terms from our set of canonical widgets~\cite{ourRepo} discussed in Section~\ref{sec:appr:ir_model} and their synonyms). 

If the heuristics alone yield a number of recommendations below the set threshold, as the next step \toolname will predict the top-1 classification of the canonical category for each interactive widget on the target app's screen. If the target-app widgets that match the expected canonical widgets in the IR Model bring the total number of matched widgets above the threshold, the process terminates and the identified widgets are presented to the developer. 
Otherwise, as the final step, the matching criteria are relaxed and the process switches from the top-1 to the top-5 classifications of each widget's canonical category.

\subsubsection{{Event Generator}}
\label{sec:appr:event_generator}

With a chosen widget, \toolname generates an executable event to trigger based on whether the widget requires user input. 
This is determined by the widget type. For example, \texttt{\small EditText}~\cite{androidEditText} is a widget type that requires an input from the user, such as entering the email address. 
In such cases, \toolname prompts the developer for text inputs, as these typically do not generalize across  apps. 
If the selected widget does not require user input, the Event Generator 
automatically executes the touch event (e.g., tap, swipe) stored in the transition of the IR Model. 

This event generation process requires minimal effort from the developer and provides the flexibility to test the same usage with different desired text inputs of the developer's choice. A test scenario is generated when the end condition is met, as discussed earlier, and each test consists of a sequence of events triggered by the Event Generator.

\subsection{\toolname's Implementation}
\label{sec:appr:implementation}

\looseness-1
\toolname is implemented in Python with 10,700 SLOC, of which the screen and widget classifiers are  stand-alone modules totaling 2,800~SLOC, and include the autoencoder model we developed.
\toolname additionally extended several research tools, including the modified V2S (500 SLOC in Python), UIED (300 SLOC in Python), and the re-implemented REMAUI (5,000 SLOC in Java).
\toolname employs the \texttt{pytransitions} library~\cite{pytransitions2021} to manipulate its state-machine IR Models, and uses the Appium testing framework~\cite{appium} for its test generation.



%

%% file: eval.tex

\section{Evaluation of \toolname}
\label{sec:eval}

\looseness-2
To demonstrate \toolname's effectiveness at generating usage tests, and its
improvement on the state of the art, we answer two research questions:

\begin{tcolorbox}
\vspace{-1mm}
\begin{itemize}

  \item[\textbf{RQ1}] How effective is \toolname at generating tests that
  exercise the desired usage?

  \item[\textbf{RQ2}] How accurate are \toolname's vision-only screen and
  widget classifiers?

  
\end{itemize}

\end{tcolorbox}
\vspace{-2mm}
\input{data}

\subsection{RQ1: \toolname's Test Quality}
\label{sec:eval:rq1}

%
%

\toolname's main goal is to generate a test for a target app that
accomplishes the usage encoded by the videos of other apps. To evaluate how
well \toolname accomplishes that goal, for each of the 18 app usages, we
randomly selected 3 apps under test (AUTs) as the target apps. (For three of the
usages, we selected only 2 apps because we were unable to extract data from
certain commercial apps due to security reasons or limitations of the
underlying used testing framework~\cite{appium}.) For
each of those apps, we use the merged IR Model \toolname learned from all the
other 17 apps to guide \toolname's test generation, aiming to demonstrate \toolname's ability to generate tests for \emph{unseen} apps. 
As discussed in Section~\ref{sec:appr:guided teset generation}, \toolname's test generation is a developer-in-the-loop process. Specifically, four of the authors served as the developers interacting with the tool during this process.
We use the first
test \toolname generates for the evaluation, resulting in a total of 51 tests across 18 app usages. (We
limit our evaluation to a single test per usage per app due to the
significant manual effort involved in the evaluation process.)

We examined each of the tests manually to consider whether it accomplished
the intended usage. Note that this judgement is objective. For example, it is
straightforward and unambiguous to determine whether the generated test
accomplishes the Sign In usage\,---\,either the tests signs into the app or it
does not. We found that 35 of the 51 tests (68.6\%) accomplished the
usage, meaning that \toolname successfully generated a correct test.

For the remaining 16 tests, we measured how similar each generated test was
to the closest human test, to evaluate whether it would potentially save
human effort in writing the test. 
The is due to the nature of \emph{usage-based} tests, as there are usually multiple correct paths of exercising the same usage. Thus, to enable fair comparison, we compare \toolname's test with the \emph{closest} human test.
We measured similarity using two metrics:
precision and recall in matching the human test's behavior. Precision
measures the fraction of the states and transitions in the generated test
that occur in the most-similar human test from the relevant videos. Recall
measures the fraction of the states and transitions in the most-similar human
test that occur in the generated test. 
The \emph{closest} human test is chosen using the precision similarity metric.

\begin{table}[t]
\centering
\caption{
We compare the 16 \toolname-generated tests that do not satisfy their
intended usage with the most-similar human test, to indicate how much work 
these tests may save a developer.}
\resizebox{\linewidth}{!}{%
\begin{tabular}{llcccc}
\toprule
  &                 & \multicolumn{2}{c}{\textbf{precision}}  & \multicolumn{2}{c}{\textbf{recall}}    \\ 
  & \textbf{Usage}  & \textbf{states} & \textbf{transitions}  & \textbf{states} & \textbf{transitions} \\ 
\midrule
U4 & Search          & 1.00      & 0.50       & 1.00      & 0.50 \\ 
U5 & Terms           & 0.71      & 0.29       & 0.63      & 0.33 \\ 
U9 & About           & 1.00      & 0.50       & 1.00      & 0.25 \\ 
U10 & Contact        & 0.75      & 0.37       & 0.72      & 0.33 \\ 
U11 & Help           & 0.67      & 0.50       & 0.67      & 0.33 \\ 
U12 & AddCart        & 0.75      & 0.59       & 0.55      & 0.37 \\ 
U13 & RemoveCart     & 1.00      & 0.69       & 0.62      & 0.42 \\ 
U14 & Address        & 0.83      & 0.69       & 1.00      & 0.75 \\ 
U15 & Filter         & 1.00      & 0.50       & 1.00      & 0.40 \\
U17 & RemoveBookmark & 1.00      & 0.75       & 0.60      & 0.50 \\ 
U18 & Textsize       & 0.00      & 0.00       & 0.00      & 0.00 \\ 
\midrule
\multicolumn{2}{l}{average} & 0.79 & 0.47 & 0.68       & 0.37 \\ 
\bottomrule
\end{tabular}%
}
\label{fig:test_results}
\vspace{-1.5em}
\end{table}

Table~\ref{fig:test_results} lists the similarity results for the 16 tests
across 11 usages that do not satisfy that usage, and the cloest human
test.
On average, 79\% of the states and 47\% of the transitions in the
generated tests is captured by the most similar human test. This means that \toolname rarely visited an incorrect state, but often triggered inputs for GUI widgets not triggered by humans.
Meanwhile, on average, the generated tests capture 68\% of
the states and 37\% of the transitions in the closest human test.
This means that \toolname was able to visit a majority of the screens seen
in the human test, but correctly exercised comparatively fewer expected GUI widgets
that trigger proper transitions.
This suggests that while the 31.4\% of the tests \toolname generates do not
fully exercise the intended usage, they may be at least partially helpful for
developers writing tests. 
Our future work will examine the effort reduction \toolname's tests produce for
developers.

\begin{tcolorbox}

RA1: We find that 69\% of \toolname's generated tests successfully
accomplish the desired usage, saving the developer from having to manually
write the test from scratch. For the remaining 31\% of the tests, we found that those
tests capture significant portions of the behavior in the most-similar test a
human would write, again, potentially saving human effort.

\end{tcolorbox}


\subsection{RQ2: \toolname's Classification Accuracy}
\label{sec:eval:rq2}


RQ2 compares \toolname's vision-only screen classification and widget
classification accuracy to the state-of-the-art S2V~\cite{li2021screen2vec}.
First, Section~\ref{sec:eval:offline_classification} evaluates \toolname's
classification independently, as a stand-alone tool. Then,
Section~\ref{sec:eval:dynamic_classification} evaluates \toolname's
classification in the context of test generation.

\subsubsection{Evaluating \toolname's Stand-Alone Classification}
\label{sec:eval:offline_classification}

\looseness-1
To evaluate \toolname's vision-only classification module as a stand-alone
technique, we use our labeled dataset from Section~\ref{sec:eval:annotation}.
We use leave-one-out cross-validation~\cite{bishop2006pattern} to evaluate
the accuracy of \toolname's screen and widget classifiers. For each of the 18
apps, we train our model on the data from all the other apps, and test on
that app.

\smallskip
\noindent
\textbf{Screen Classification:} 
We evaluate three variants of
\toolname's screen classifier. The first, the standard \toolname as introduced in
Section~\ref{sec:appr}, and two other classifiers that use only
\toolname's autoencoder (AE) model and classify with two widely-adopted
methods KNN~\cite{altman1992KNN} (AE + KNN) and MLP~\cite{gardner1998MLP} (AE
+ MLP), respectively. As \toolname's AE model only encodes the screen's
visual features, the results aim to demonstrate the impact of
visual-only information on the classification tasks. We did not evaluate the
text-only model since it contains app-specific noises (e.g., news content,
product description) that do not generalize, and text-only information has
already been shown insufficient for classification tasks~\cite{hu2018appflow}.

\looseness-1
To compare \toolname with S2V, we adapt S2V to learn from the information
on the screen images only, and obtain the UI layout
information~\cite{androidUILayouts} that S2V requires as its input. To do so,
we reverse engineered app screen images using REMAUI~\cite{nguyen2015REMAUI},
a research tool to convert app screen image into its corresponding UI
layout~\cite{androidUILayouts}. This is the same process \toolname uses
(recall Section~\ref{sec:appr}), aiming to ensure that S2V and \toolname
learn from the same raw information on the app screen. We then apply KNN and
MLP to S2V's screen embeddings, resulting in two S2V's variants (S2V + KNN,
S2V + MLP).

\begin{figure}[t]
  \includegraphics[width=.48\columnwidth]{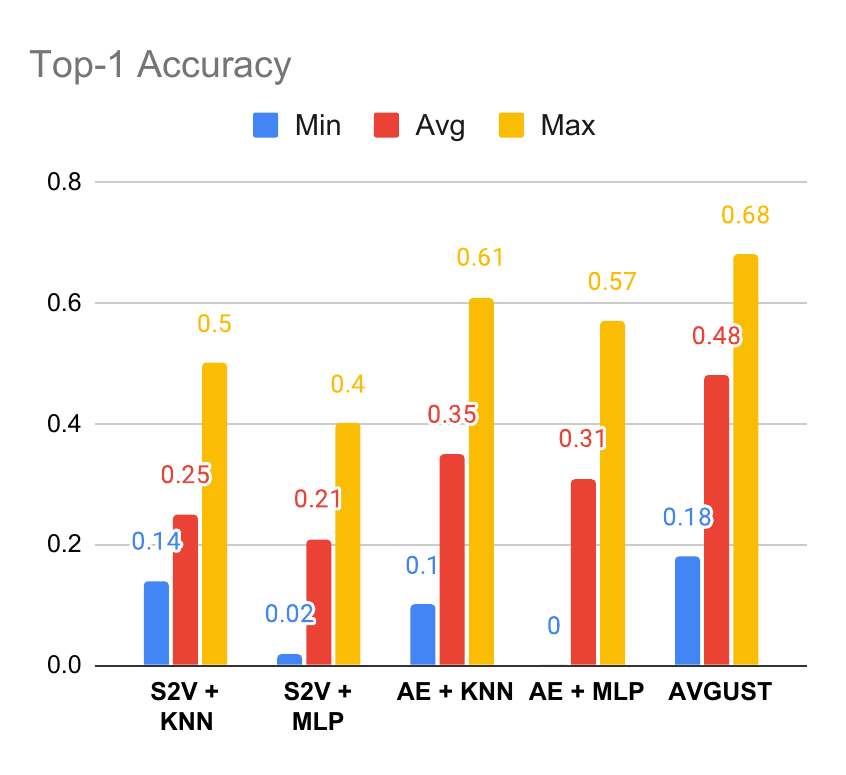} \hfill 
  \includegraphics[width=.48\columnwidth]{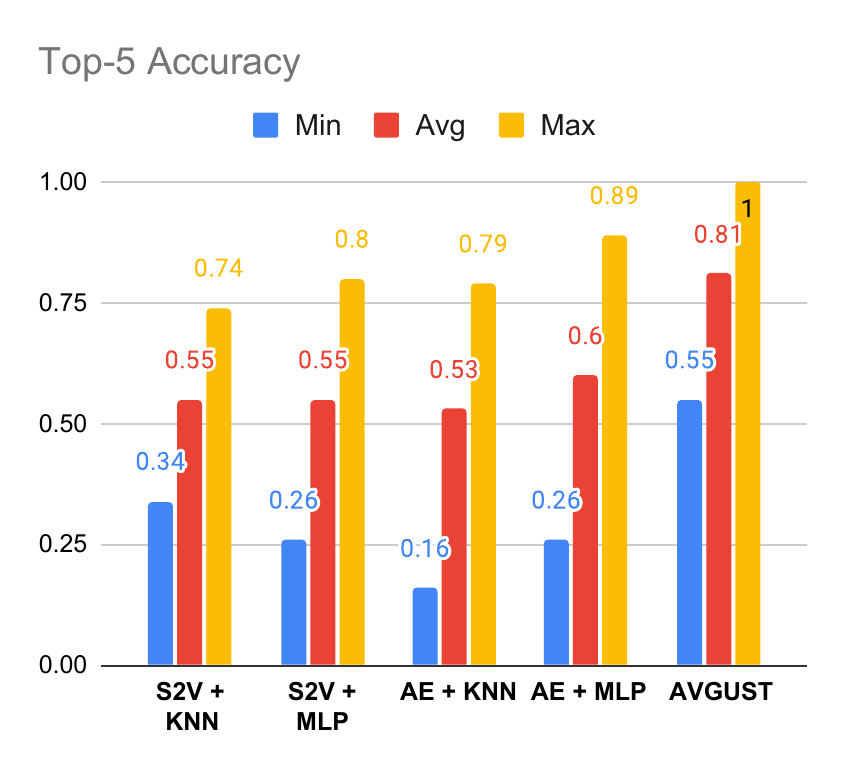}
\vspace{-1em}
  \caption{\toolname's vision-only screen classification outperforms
  SV2~\cite{li2021screen2vec} and the other \toolname classifier variants in
  both top-1 and top-5 accuracy.}
  \label{fig:offline_screen}
\vspace{-1.5em}
\end{figure}

Figure~\ref{fig:offline_screen} shows that \toolname's classifier
consistently outperforms all versions of SV2 and the other \toolname
classifier variants in both top-1 and top-5 accuracy. \toolname's composite
classifier that uses both visual and textual screen features outperforms both
autoencoder-only variants by more than 20\%, suggesting that although visual
features are important in encoding a UI screen, adding textual features
significantly improves the quality of the generated embeddings and results in
higher classification accuracy.

While using S2V's screen embeddings is effective for down-stream tasks when
the dynamic UI layout information is available~\cite{li2021screen2vec}, we
were unable to achieve high classification accuracy using the pre-trained S2V
model by reverse engineering app screen images into S2V's required format.
This suggests that the existing pre-trained models cannot be used for
vision-only tasks effectively.

\smallskip
\noindent 
\textbf{Widget Classification:}
To compare \toolname's widget classification with
S2V~\cite{li2021screen2vec}, we studied S2V's implementation and isolated its
underlying model that encodes the widget's information. We then applied both
KNN and MLP to S2V's widget embeddings.

\begin{figure}[t]
  \includegraphics[width=0.49\columnwidth]{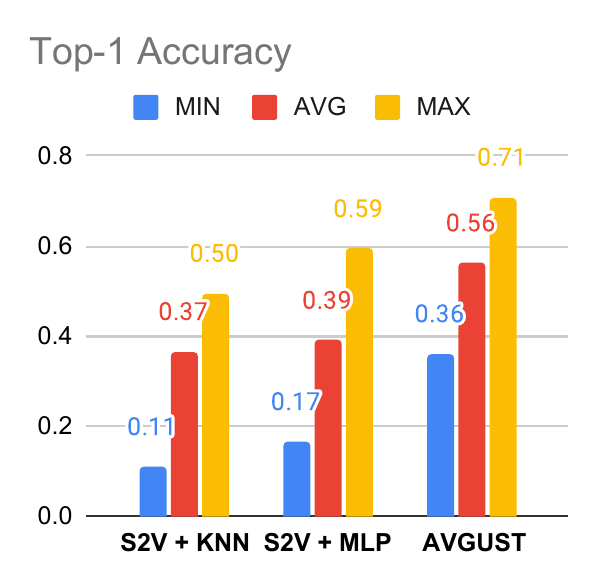} \hfill 
  \includegraphics[width=0.49\columnwidth]{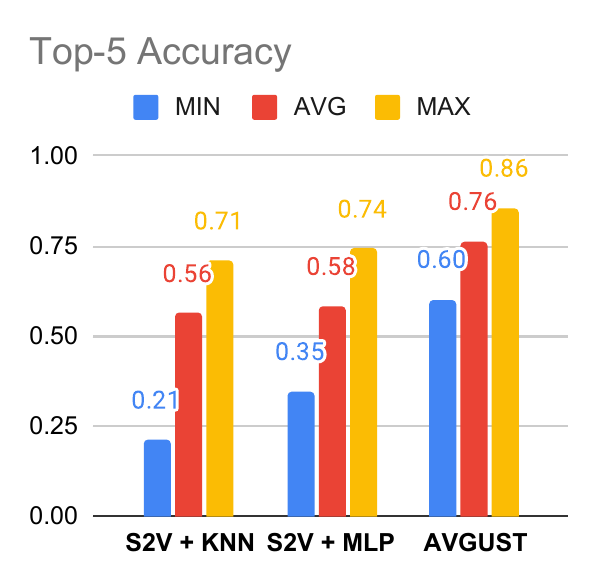}
\vspace{-1em}
  \caption{\toolname's vision-only widget classification consistently
  outperforms S2V~\cite{li2021screen2vec}.}
  \label{fig:offline_widget}
\vspace{-1.25em}


\end{figure}

Figure~\ref{fig:offline_widget} shows that \toolname's widget classifier
outperforms both S2V variants that use the pre-trained UI widget encoder for
two reasons. First, S2V's UI widget encoder only uses a widget's textual
information and class type, whereas \toolname's widget classifier takes into
account many other widget features, such as its location on the screen and
visual features. Second, S2V's UI widget encoder is trained using the textual
information available on dynamically extracted UI layout, which is not
available for the widgets in \toolname's vision-only classification task.

\subsubsection{Evaluating \toolname's Classification for Test Generation}
\label{sec:eval:dynamic_classification}

We next evaluate \toolname's classification accuracy in the context of test
generation. During the test generation phase in Section~\ref{sec:eval:rq1},
we recorded \toolname's Top-1 and Top-5 recommendations at each step, and
 evaluate the accuracy of those recommendations.

\smallskip
\noindent 
\textbf{Evaluation \toolname's Screen Classification:}
Besides \toolname's built-in screen classifier introduced in
Section~\ref{sec:appr}, we further implemented 5 variants that incorporate
app's runtime information, aiming to get insights on whether runtime
information can improve the classification's accuracy. This is inspired by
S2V~\cite{li2021screen2vec}, which relies on the screen's runtime UI layout
information~\cite{androidUILayouts}, such as the \emph{Activity
name}~\cite{AndroidActivity} and \emph{content description} of the widgets on
the screen~\cite{Androidaccessibility}. As \toolname's test generation phase
interacts with the target app at runtime, \toolname can crawl the UI for this
layout information. We thus relaxed the vision-only constraint, and modified
\toolname to use this dynamic UI layout information. We term the modified
version \toolname-Dynamic. Note that the fundamental difference between
\toolname-Dynamic and S2V is the training phase. \toolname-Dynamic still uses
vision-only information (screen images) to train its models, while S2V
requires the dynamic UI layout information in the training data. In practice,
as classification tasks usually require a large amount of training data,
\toolname-Dynamic makes the training process significantly easier by only
requiring screen images, while S2V requires crawling the UI
layout information at app runtime for every app screen in the training set.

\begin{figure}[t]

  \includegraphics[width=0.49\columnwidth]{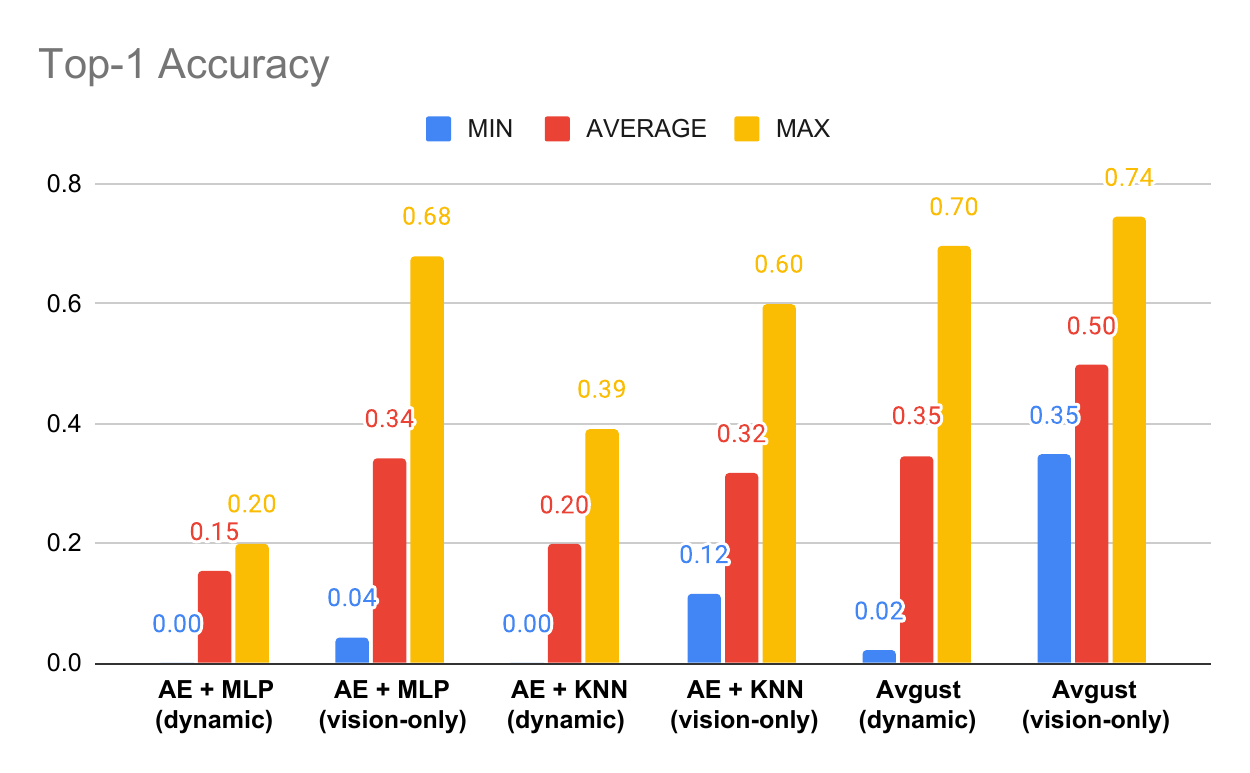}~\includegraphics[width=0.49\columnwidth]{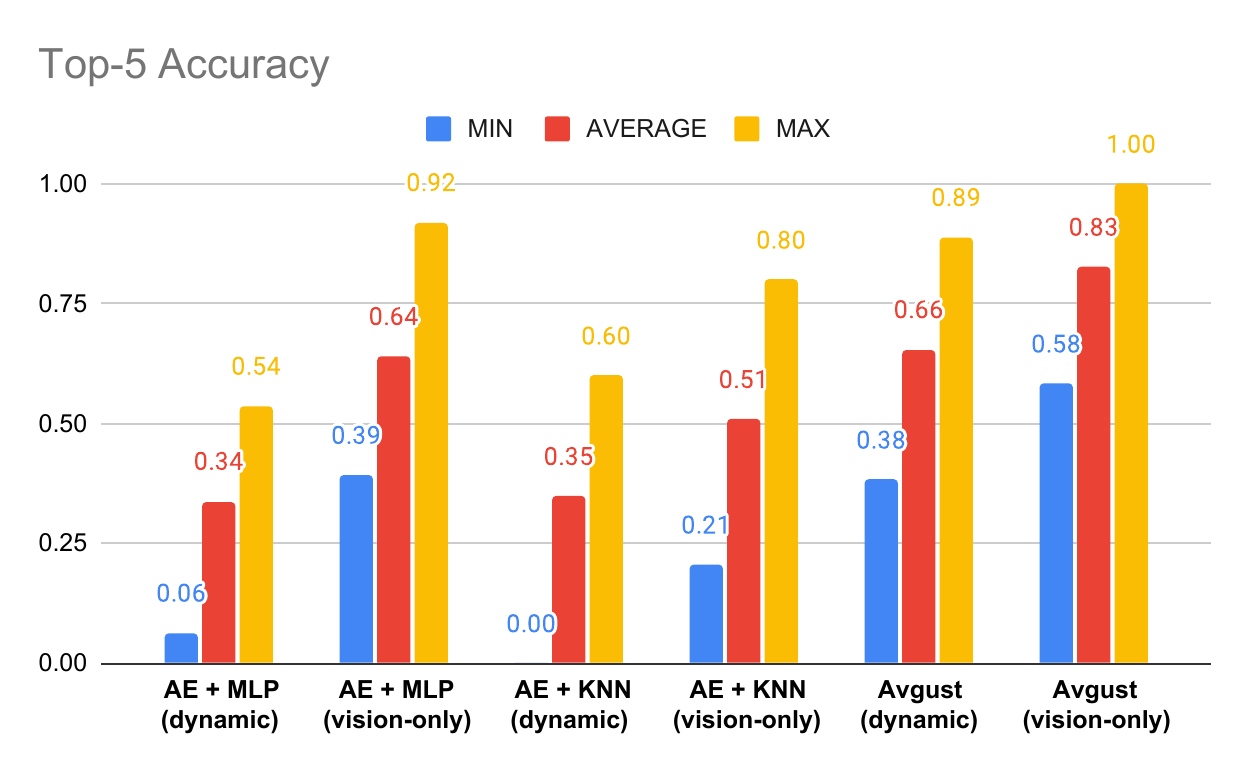}
\vspace{-1em}
  \caption{The accuracy of \toolname's vision-only screen classifier
  variations with vision-only and dynamic input data.}
  \label{fig:dynamic_screen}

\vspace{-1.5em}
\end{figure}

Figure~\ref{fig:dynamic_screen} shows \toolname's screen classifier variants'
accuracies during the test generation phase. Comparing
Figures~\ref{fig:dynamic_screen}~and~\ref{fig:offline_screen}, we observe
that in both cases, \toolname always outperforms the two autoencoder-only
variants.
All the classifier variants are more
accurate when using the vision-only data, compared to also using dynamic app
information captured during runtime. While this might seem
counterintuitive, one possible explanation is that the features extracted
from the vision-only input data are similar to the data \toolname's built-in
classifier were trained with, whereas
the dynamically obtained information might expose much more textual
information (e.g., content description) that is not consistent with the OCR-based textual information used in the training phase.

\smallskip
\noindent 
\textbf{Evaluation \toolname's Widget Classification:}
To evaluate \toolname's widget classifier during test generation, we assess
whether \toolname can faithfully recommend widgets to match the transitions
suggested by its IR Model. We recorded the next transitions suggested by the
IR Model at each step of the test generation (recall
Section~\ref{sec:appr:guided teset generation}), as well as the crops of
\toolname's recommended widgets. Three annotators then manually inspected the
cropped widgets and determined whether their canonical categories match one of the
suggested transitions. In total, over all the generated tests, \toolname's
widget classifier correctly recommended widgets 77.4\% of the time (175 out
of 226 steps).
\begin{tcolorbox}

RA2: \toolname's classifiers consistently outperform the state-of-the-art S2V
tool. We find that using textual and visual features together improves
accuracy, but adding runtime features decreases accuracy, perhaps because
these features are too different from the ones used to train \toolname's vision-only
models.

\end{tcolorbox}
\vspace{-0.5em}

%% file: data.tex

\begin{table}[b!]
\centering
\caption{The 18 usages used in \toolname's evaluation.}
\resizebox{\linewidth}{!}{%
\begin{tabular}{clll}
\toprule
\textbf{Usage ID} & \textbf{Test Case Name} & \textbf{Tested Functionalities}                 & \textbf{\#Videos} \\ \hline
U1  & Sign In      & provide username and password to sign in              & 21\\ 
U2  & Sign Up      & provide required information to sign up               & 76\\ 
U3  & Search       & use search bar to search a product/news               & 29\\ 
U4  & Detail       & find and open details of the first search result item & 17\\ 
U5  & Category     & find first category and open browsing page for it     & 27\\ 
U6  & About        & find and open about information of the app            & 15\\ 
U7  & Account      & find and open account management page                 & 18\\ 
U8  & Help         & find and open help page of the app                    & 17\\ 
U9  & Menu         & find and open primary app menu                        & 12\\ 
U10 & Contact      & find and open contact page of the app                 & 16\\ 
U11 & Terms        & find and open legal information of the app            & 20\\ 
U12 & Add Cart     & add the first search result item to cart              & 13\\ 
U13 & Remove Cart  & open cart and remove the first item from cart         & 10\\ 
U14 & Address      & add a new address to the account                      & 11\\ 
U15 & Filter       & filter/sort search results                            & 14\\ 
U16 & Add Bookmark & add first search result item to the bookmark          & 15\\ 
U17 & Remove Bookmark         & open the bookmark and remove first item from it & 20                  \\ 
U18 & Textsize     & change text size                                      &  23\\ 
\bottomrule
\end{tabular}%
}
\label{fig:usage_data}
\end{table}

\subsection{Evaluation Context}
\label{sec:eval:setup}

To evaluate \toolname as a whole, app videos are needed as its input. To
collect these videos, we relied on the apps and usages defined by the FrUITeR
benchmark~\cite{yixue_zhao_fruiter_2020}, which contains 20 popular apps and
18 most-common app usages. We then designed a user study to collect screen
recordings of these app usages, which resulted in the collection of 374 videos.
To evaluate \toolname's image classification component, we developed a
semi-automated pipeline to annotate the video frames of screen recordings
collected by users with ground-truth canonical categories for both screens
and GUI widgets. This process is detailed in
Section~\ref{sec:eval:annotation}.

\subsubsection{Video Collection}
\label{sec:eval:video}





We designed a large-scale user study to collect videos of app usages from participants. We recruited and assigned the 18 usages and 20 apps to 61 computer science students in a Master's level course at the authors' institution, and asked them to record videos of themselves exercising scenarios that triggered the features associated with the usages. We assigned usages such that each student was assigned 2 applications, each with 2 different usages, for a combination of 4 app-usage pairs. We balanced the assigned apps and usages evenly across participants. We then asked them to collect two screen-recording videos for each app-usage pair, for a  \textit{potential} total of 8 videos per participant. We asked for \textit{two} videos per app-usage pair in order to capture different ways of exercising a given feature (e.g., adding an item to a shopping cart by searching vs. by browsing categories). However, if a participant deemed that there were not two distinct scenarios for exercising a given feature, they were allowed to provide only one usage.

This study was conducted remotely due to COVID-19, and participants were given detailed instructions for installing and setting up an Android emulator (Nexus 5X, API24), the \texttt{\small .apk} files required to install their assigned apps, and a short textual description of the assigned use cases (illustrated in Table~\ref{fig:usage_data}). Additionally, we provided a small desktop application that allowed participants to record the screens and a usage trace of their scenarios. This application makes use of the \texttt{\small adb} \texttt{\small screenrecord} and Linux \texttt{\small getevent} command line utilities. We provide these instructions, the app \texttt{\small .apk} files, usage descriptions, device recording tool, and anonymized collected data in our online appendix~\cite{ourRepo}. This study was approved by the Institutional Review Board (IRB) at the authors' university (IRBNet 1666261-1).

This data collection process spanned two semesters, and in total, 31 of the originally recruited 61 students completed the study, some with partial data, hence the imbalance of videos across apps and usages shown in Tables~\ref{fig:usage_data}~and~\ref{fig:apps_data}. In the end, we obtained a dataset of 374 screen recordings of 18 usages from 18 of the 20 apps (shown in Table~\ref{fig:apps_data}) from the {FrUITeR} benchmark~\cite{yixue_zhao_fruiter_2020} discussed earlier.

\begin{table}[b!]
\vspace{-1.5em}
\footnotesize
\caption{The 18 subject apps used in \toolname's evaluation.}
\centering
\begin{tabular}{clcl}
\toprule
\textbf{App ID} & \textbf{App Name} & \textbf{\#Downloads (mil)} & \textbf{\#Videos} \\ 
\phantom{0}A1   & AliExpress       & 100\phantom{.000}                 &    ~~27 \\ 
\phantom{0}A2   & Ebay              & 100\phantom{.000}                 &      ~~15             \\ 
\phantom{0}A3   & Etsy              & \phantom{0}10\phantom{.000}       &      ~25             \\ 
\phantom{0}A4   & Dailyhunt         & \phantom{00}1.7\phantom{00}                 &   ~\phantom{0}8                \\ 
\phantom{0}A5   & Geek              & \phantom{0}10\phantom{.000}                  &  ~17                 \\ 
\phantom{0}A6   & Groupon           & \phantom{0}50\phantom{.000}                  & ~ 53                 \\ 
\phantom{0}A7   & Home              & \phantom{0}10\phantom{.000}                  & ~  53                \\ 
\phantom{0}A8   & 6PM               & \phantom{00}0.5\phantom{00}                 &  ~ 25                \\ 
\phantom{0}A9   & Wish              & 100\phantom{.000}                 &      ~     44        \\ 
A10             & The Guardian      & \phantom{00}5\phantom{.000}                   &~  \phantom{0}8                 \\ 
A11             & ABC News          & \phantom{00}5\phantom{.000}                   &  ~ 18                \\ 
A12             & USA Today         & \phantom{00}5\phantom{.000}                   &  ~ 26                \\ 
A13             & Zappos            & \phantom{00}0.054                  &      ~   11          \\ 
A14             & BuzzFeed          & \phantom{00}5\phantom{.000}                   & ~ \phantom{0}8                 \\ 
A15             & Fox News          & \phantom{0}10\phantom{.000}                  & ~ 11                 \\ 
A16             & BBC News          & \phantom{0}10\phantom{.000}                  &  ~ \phantom{0}6                \\ 
A17             & Reuters           & \phantom{00}1\phantom{.000}                   & ~ 12                 \\ 
A18             & News Break        & \phantom{00}0.322                 &      ~   \phantom{0}7          \\ 
\bottomrule
\end{tabular}%
\label{fig:apps_data}
\end{table}

\subsubsection{Ground-Truth Annotation}
\label{sec:eval:annotation}

Recall from Section~\ref{sec:appr} that \toolname's classification 
involves (1)~the screen classifier that maps an app screen to an
abstract screen IR category, and (2)~the widget classifier that maps a cropped
widget image to an app-independent {canonical widget category}. To establish
the ground-truth labels (i.e., the correct categories needed to generate
\toolname's IR Models), we developed a pipeline to import pairs of
screen-widget images into Label Studio~\cite{labelstudio}, and trained four human
annotators to label the data.
Specifically, the screen-widget image pairs are sourced from the \emph{GUI
Event Frames} that \toolname converted during its \textit{Video Analysis}
phase (recall Figure~\ref{fig:design}).

To establish our ground truth categorizations, we provided detailed
instructions to and trained the four annotators to label the data based on the 37
screen and 74 widget {canonical categories} we defined (recall Section~\ref{sec:appr:ir_model}). 
Each image is labeled by at least two annotators, and discrepancies are resolved
by negotiated agreement~\cite{Charmaz:groundedtheory} with the annotators and
one author.

This data collection process was time-consuming and intensive, spanning
$\sim\!\!8$ person-months of effort. At the end of the process, we derived a
comprehensive labeled dataset containing 2,478 ground-truth labels for
screens and 2,434 labels for widgets across 18 apps. Given these labels,
\toolname was able to automatically generate the IR Models for the usages
needed for our evaluation. Our labeled dataset, as well as the annotation
pipeline we developed can be easily reused or extended by future work in this
area~\cite{ourRepo}.



%

%

%% file: related-work.tex
\section{Related Work}
\label{sec:bg:related_work}
%
%
%
%

\noindent\textbf{Automated Input Generation for Mobile Apps:} Existing automated test generation techniques share a complementary objective to ours: they mainly focus on generating tests to maximize code coverage and detect crashes, as opposed to generating \emph{usage-based} tests to test a certain functionality. The large body of existing work includes model-based testing~\cite{su2017guided, amalfitano2012using, gu2019practicalAPE, dong2020time, lukic2020airmochi, salihu2019amoga, amalfitano2014mobiguitar, su2020automatedGenie, lukic2020remote}, random testing~\cite{monkeytool, machiry2013dynodroid,YazdaniBanafsheDaragh2021DeepGUI}, and systematic testing~\cite{mao2016sapienz, mahmood2014evodroid, adamsen2015systematic, azim2013targeted}.
Recently, Su et al. proposed Genie~\cite{suOOPSLA2021Genie}, which is the first automated testing technique to detecting non-crashing functional bugs in Android apps. However, Genie is a random-based fuzzing technique, thus does not generate usage-based tests.
Besides the differences in testing objectives,  \toolname's model is app-independent (representing usage scenarios learned from different apps) and  is derived purely from visual data, which differs from existing model-based testing techniques.  Furthermore, in comparison to recent human-in-the-loop techniques, e.g. NaviDroid~\cite{liu2022guided}, \toolname's model and recommendations differ by providing suggestions for GUI actions that fulfill a given use case, as opposed to uncovering unexplored areas of an app.

\noindent\textbf{Test Reuse in Mobile Apps:} The area of research that most closely aligns with usage-based test generation is the work on \emph{UI test reuse}, which has been steadily growing over the past few years~\cite{yixue_zhao_fruiter_2020, behrang2018test, lin2019craftdroid, behrang2019atm, mariani2021evolutionary, mariani2021semantic, qin2019testmig, Talebipour2021MAPIT}.  These techniques can transfer an existing usage-based test from a \emph{source app} to its equivalent test of a \emph{target app} that shares the same functionality, but cannot  generate usage-based tests from scratch. Furthermore, as discussed in Section~\ref{sec:intro}, existing test-reuse techniques have three important limitations that we directly address in this paper.

\noindent\textbf{Learning Patterns from Crowdsourced Tests:} Similar to our objective, another line of work aims to learn patterns from crowdsourced tests for automated test generation. However, while such techniques learn from crowdsourced data, their test objectives are to increase coverage or fault-finding ability as opposed to generating usage-based tests. For example, Replica \cite{wang2020improvingReplica} compares existing in-house tests with the user traces in the field, and generates new tests to mimic field traces that are not covered by the in-house tests. Replica relies on pre-existing in-house tests that may not be available, as well as app instrumentation. Ermuth et al. proposed an approach to generate ``macro events'' that group multiple low-level events into logic steps performed by real users~\cite{ermuth2016monkeysee}, such as filling and submitting a form. However, these macro events are recurring patterns across \emph{all} the user traces collected when exercising the \emph{entire} app, thus do not capture fine-grained user behaviors exercised in specific usages. Similarly, MonkeyLab and Polariz~\cite{linares2015miningMonkeyLab,Mao:ASE17} mines users' event traces to generate combinations of low-level events representing natural scenarios (similar to ``macro events''), as well as untested corner cases (similar to Replica's objective). 
ComboDroid~\cite{wang2020combodroid} aims to reach complex app functionality by combining independent short ``use cases'', such as toggling a setting, or switching to a different screen.  Humanoid~\cite{li2019humanoid} leverages a deep neural network model to learn input actions  based on real-user traces.  However, the generated tests from all the work mentioned above again focus on maximizing the code coverage, but do not aim to generate tests of specific usages.

\noindent\textbf{Specification-based Testing for Mobile Apps:} Finally, this type of testing aims to generate tests that cover specific functionalities (similar to our definition of \emph{usages}), guided by manually-written specifications. For example, FARLEAD-Android~\cite{koroglu2021functionalFARLEAD} requires the developer to provide UI test scenarios written in Gherkin~\cite{Gherkin} in order to generate tests using reinforcement learning. Similarly, AppFlow~\cite{hu2018appflow} requires the developer to first create a test library that covers the common functionalities in a certain app category (e.g., shopping apps) using a Gherkin-based language that AppFlow defines. AppFlow then synthesizes app-specific tests according to the test library. Augusto~\cite{mariani2018augusto} uses GUI ripping to explore popular app-independent functionalities (referred to as ``AIFs''), and generates functional tests accordingly. However, developers have to manually define UI patterns and Alloy semantic model~\cite{jackson2002alloy} to describe the AIFs. 
 \toolname attempts to advance upon such work by simplifying the specification process by relying purely on videos that specify desired test behaviors. 

%% file: conclusion.tex
\vspace{-2mm}
\section{Contributions}
\label{sec:conclusion}

We have presented \toolname, a method for generating usage-based tests for
the Android platform. By targeting usage-based tests, \toolname solves what
mobile developers identify as a major need~\cite{linares2017developers} but
that the state of the art has failed to address~\cite{linares2017developers,
Kochar2015Study, haas2021manual}. \toolname uses user-generated videos of app
usages to learn a model of a usage, and then applies that model to a new
target app to generate tests. Evaluating on 374 videos of common uses of 18
popular apps, we show that 69\% of the tests \toolname generates successfully
execute the desired usage, that the remaining generated tests have potential
for reducing developer effort in writing tests, and that \toolname's
classifiers outperform the state of the art. Our work suggests a promising
direction of research into usage-based test generation, and outlines
outstanding problems in classification accuracy that future research should
address.

%
\vspace{-2mm}
\section*{acknowledgement}
This work is supported by the U.S.\ National Science Foundation under grant
no.\ CCF-1717963, CCF-1763423, CNS-1823354, CCF-1955853, and CCF-2030859 (to
the Computing Research Association for the CIFellows Project), as well as the
U.S.\ Office of Naval Research under grant N00014-17-1-2896. Additionally, we
would like to thank Arthur Wu for his help on data collection and annotation.